\documentclass[a4paper,11pt]{article}
\usepackage[latin1]{inputenc}
\usepackage[english]{babel}
\usepackage{graphicx}
\usepackage{latexsym}
\begin{document}
\title{Bianchi type $IX$ asymptotical behaviours with a massive scalar field: chaos strikes back}
\author{Stéphane Fay\footnote{steph.fay@wanadoo.fr} and Thierry Lehner\footnote{thierry.lehner@obspm.fr}\\
Laboratoire Univers et Théories (LUTH), UMR 8102\\
Observatoire de Paris, F-92195 Meudon Cedex\\
France}
\maketitle
\begin{center}
\begin{abstract}
We use numerical integrations to study the asymptotical behaviour of a homogeneous but anisotropic Bianchi type $IX$ model in General Relativity with a massive scalar field. As it is well known, for a Brans-Dicke theory, the asymptotical behaviour of the metric functions is ruled only by the Brans-Dicke coupling constant $\omega_0$ with respect to the value $-3/2$. In this paper we examine if such a condition still exists with a massive scalar field. We also show that, contrary to what occurs for a massless scalar field, the singularity oscillatory approach may exist in presence of a massive scalar field having a positive energy density.
\end{abstract}
\end{center}
\section{Introduction}\label{s0}
Historically, scalar fields in cosmology were introduced by Dirac \cite{Dir37} and Kaluza-Klein \cite{Kal21} works and then from Jordan, Brans and Dicke \cite{BraDic61,Bra97} first scalar-tensor theory. Today they find new justification through particles physics theories. Hence the Higgs mechanism explaining the mass of the particles is a massive scalar field \cite{Zel86}. The supersymmetry concept which assumes a new symmetry between the fermions and the bosons, predicts that several scalar fields should exist. In cosmology, the recent acceleration of the Universe expansion can be explained by a quintessent scalar field. It is also one of the best mechanism to produce inflation at early times. It could even mimic the galactic dark matter \cite{MatGuzUre99, Fay03A} responsible for the rotation curves flattening.\\
What about the geometrical framework of this paper? Most of times, an isotropic and homogeneous Universe, and thus a Friedmann-Lemaître-Robertson-Walker(FLRW) metric, isI assumed. This is nicely confirmed by WMAP \cite{Spe03} observations. However the geometry could be different at early times. A way to generalise it is to consider a homogenous but anisotropic cosmological model of Bianchi. Nine Bianchi models exist and the Bianchi type IX one is probably the most interesting. Following the Belinskii-Khalatnikov-Lifshitz (BKL) conjecture \cite{BelKhaLif82,BelKhaLif70}, its oscillatory singularity approach would be shared by the most general inhomogeneous models \cite{UggElsWaiEll03}.\\
In this paper, we examine the metric functions asymptotical behaviour of the Bianchi type $IX$ model in General Relativity with a massive scalar field. A similar study was performed in \cite{TopUst99} with a specific form of potential. With only a perfect fluid, the metric functions oscillate in the vicinity of the singularity, adopting the so-called Mixmaster behaviour. With a massless scalar field, V. A. Belinskii and I. M. Khalatnikov\cite{BelKha73} shown that the oscillations were destroyed when the Universe collapse to singularity. In fact their results applied for a massless scalar field having a positive energy density or, equivalently, respecting the weak energy condition\footnote{The weak energy condition is that the sum of the scalar field pressure $p_\phi$ and density $\rho_\phi$ be positive. For General Relativity with a massless scalar field for which $p_\phi=\rho_\phi$, equation of state of a stiff fluid, it is equivalent to write that the scalar field energy density is positive.}. However, this is not always the case, in particular if we consider the Brans-Dicke theory. Then in the Einstein frame, the behaviour of the metric functions strongly depends on the control parameter $\omega$ called the Brans-Dicke coupling constant. As it is well known, when $\omega>-3/2$, the singularity approach is monotonic. The metric functions adopt a kasnerian behaviour, decreasing as some power of the proper time. However, when $\omega<-3/2$, the metric functions oscillate forever and there is no singularity. Heuristic argument about these facts using the BKL approximation are shown in the appendix whereas some considerations about chaos may be found in \cite{LehMen03} or in section \ref{s4}. They are valid whatever the massless scalar field theory, i.e. even if $\omega=\omega(\phi)$ instead of being a constant. Note also that General Relativity with a massless scalar field is equivalent to General Relativity with a stiff fluid having a positive or negative energy density. Hence, dynamical properties of the Universe (that is singularity, oscillations, monotonic behaviour, etc) in such a theory do not depend on the initial conditions. Moreover, a massless scalar field seems to kill the singularity oscillatory approach which is one of the main interest of the Bianchi type $IX$ model.\\
One of the goal of this paper will be to know what happens in presence of a massive scalar field. Considering several scalar field potentials, we will show that dynamical properties of the Universe now depend on the initial conditions. However some properties are still ruled only by some control parameters (i.e. the Brans-Dicke coupling constant and the constant appearing in the potential of the scalar field). Using a toy model, a second goal will be to show that a massive scalar field having a positive energy density does not always prevent from having a singularity oscillatory approach.\\
The organisation of the paper is as follows. In section \ref{s1}, we write the field equations. In section \ref{s2}, we numerically analyse the asymptotical behaviours of three theories with respect to their control parameters. In section \ref{s3}, we try to answer the question: is an oscillatory approach of the singularity always killed by a massive scalar field having a positive energy density? We will produce a toy model that shows that the answer to this question is negative. In section \ref{s4}, a Hamiltonian approach is used to detect the presence of chaos for the theories of section \ref{s2}. We conclude in section \ref{s6}.
\section{Field equations}\label{s1}
The action of General Relativity with a massive scalar field is:
\begin{equation} \label{lagrangien}
S=\int (R-(\omega+3/2)\phi^{-2}\phi_{,\mu}\phi^{,\mu}-U)\sqrt{-g}
\end{equation}
$\omega$ is the Brans-Dicke function describing the coupling between the metric and the scalar field $\phi$. $U$ is the potential describing the self coupling of the scalar field.\\
The diagonal metric of the Bianchi type $IX$ model is given by:
\begin{equation} \label{metrique}
ds^2=-dt^2+g_{\mu\mu}(\omega^\mu)^2=-e^{2\alpha+2\beta+2\gamma}d\tau ^2+e^{2\alpha}(\omega^1)^2+e^{2\beta}(\omega^2)^2+e^{2\gamma}(\omega^3)^2
\end{equation}
The $\omega^i$ are the 1-forms defining the Bianchi type $IX$ model. The field equations are got by varying the action with respect to the metric functions and the scalar field. In the $\tau$ time such as $dt=e^{\alpha+\beta+\gamma}d\tau$, they write:
\begin{eqnarray}
\ddot{\alpha}&=&\frac{1}{2}\left[(e^{2\beta}-e^{2\gamma})^2-e^{4\alpha}\right]+\frac{1}{2}e^{2(\alpha+\beta+\gamma)}U\label{eq1}\\
\ddot{\beta}&=&\frac{1}{2}\left[(e^{2\alpha}-e^{2\gamma})^2-e^{4\beta}\right]+\frac{1}{2}e^{2(\alpha+\beta+\gamma)}U\label{eq2}\\
\ddot{\gamma}&=&\frac{1}{2}\left[(e^{2\alpha}-e^{2\beta})^2-e^{4\gamma}\right]+\frac{1}{2}e^{2(\alpha+\beta+\gamma)}U\label{eq3}\\
\dot{\alpha} \dot{\beta} +\dot{\alpha} \dot{\gamma} +\dot{\beta} \dot{\gamma} &=&\frac{1}{4}\left[e^{4\alpha}+e^{4\beta}+e^{4\gamma}-2(e^{2(\alpha+\beta)}+e^{2(\alpha+\gamma)}+e^{2(\beta+\gamma)})\right]\nonumber \\
&&+\frac{1}{2}e^{2(\alpha+\beta+\gamma)}U+\frac{1}{2}\frac{\dot{\phi}^2}{\phi^{2}}(\omega+\frac{3}{2})\label{eq4}\\
0&=&\dot{\phi}^{2}\left[-\frac{\omega_{\phi}}{\phi^2}+\frac{3+2\omega}{\phi^{3}}\right]-\frac{3+2\omega}{\phi^2}\ddot\phi-e^{2(\alpha+\beta+\gamma)}U_\phi\label{eq5}\\\nonumber
\end{eqnarray}
A dot means a derivative with respect to $\tau$ whereas $U_\phi$ is the potential derivative with respect to the scalar field. In this paper, we will choose $\omega=\omega_0$ where $\omega_0$ is a constant. In the next section, we integrate numerically these equations.
\section{Numerical study of the Bianchi type $IX$ model}\label{s2}
Without the potential and $\omega=\omega_0$, the above action is the Brans-Dicke action in the Einstein frame \cite{Dic62}. Metric functions asymptotical behaviour (that is oscillation or monotonicity) is then ruled by the value of $\omega_0$ with respect to $-3/2$ as explained in the introduction. However we have to note that when $\omega<-3/2$ there are some known problems. Hence, a scalar field negative kinetic energy leads to an instability of the empty Minkowski spacetime or some ghosts and tachyons propagation in linearised field equations.\\
In this section, we want to know if some similar results exist when we consider a massive scalar field: does the metric functions asymptotical behaviour continue to depend only on some control parameters or do they also depend on the initial conditions?\\
We will consider three potentials that are widely used to explain dark energy or inflation:
\begin{itemize}
\item $U=2\Lambda$, i.e. a cosmological constant which could model vacuum energy.
\item $U=\phi^k$, i.e. a power law potential whose Ratra-Peebles\cite{RatPee88} one is a subcase with $k<0$. It appears in models of supersymmetric QCD\cite{Bin99}.
\item $U=e^{k\phi}$, i.e. an exponential law potential. It can not model quintessence but is the outcome of compactification of higher-dimensional theories.
\end{itemize}
In the Brans-Dicke frame, the two first theories may be cast into the Brans-Dicke theory with a power law potential. The third one has been studied in  \cite{ColIbaHoo97,KitMae92,Fay01,Fay03} from the Bianchi models isotropisation point of view. In \cite{TopUst99}, the case with $U=m^2\phi^2$ was studied.\\
Since we want to find what rules the metric functions behaviour (the control parameters? the initial conditions?), we need to identify what are these behaviours. During the numerical simulations (see below), we identified several of them, sometimes different from those encountered in the Brans-Dicke theory:
\begin{figure}[h]
\centering
\includegraphics[width=8cm]{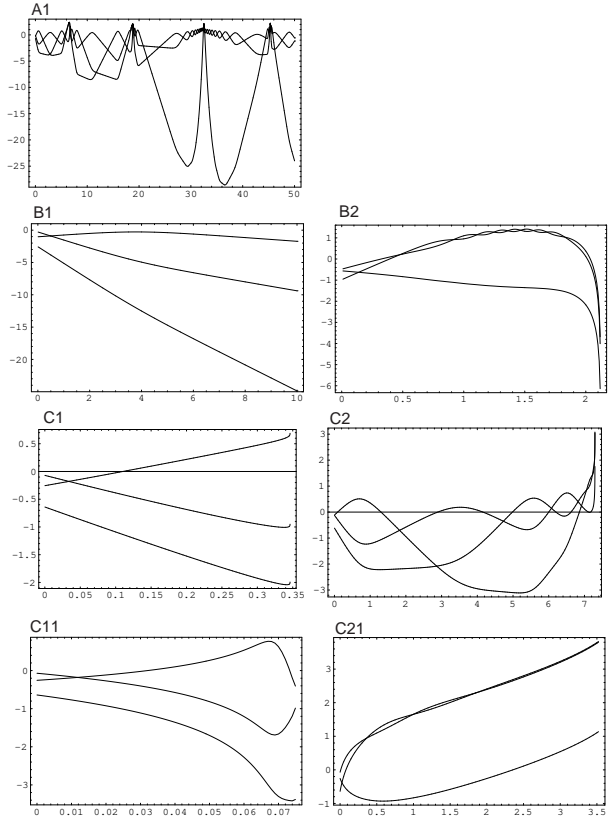}
\caption{\label{fig0}\scriptsize{The three metric functions asymptotical behaviours encountered during numerical simulations: oscillation($A_1$), singularity monotonic approach(Kasnerian for $B_1$, non kasnerian for $B_2$), expansion (anisotropic expansion for $C_1$ and $C_{11}$ respectively in the $\tau$ and $t$ times, isotropic expansion for $C_2$ and $C_{21}$ respectively in the $\tau$ and $t$ times).}}
\end{figure}
\begin{itemize}
\item The first one, represented on graph $A_1$, is such that the oscillations never end. It is what occurs for the Brans-Dicke theory in the Einstein frame when the massless scalar field energy density is negative ($\omega<-3/2$) or equivalently in General Relativity with a stiff perfect fluid having a negative energy density. Contrary to General Relativity in the empty or with a perfect fluid of dust or radiation, these oscillations do not lead to a singularity: the Universe 3-volume does not go to zero in the $\tau$ or $t$ time. Moreover, the oscillatory behaviour can not be continuously approximated by the BKL approximation\footnote{See appendix for a description of this approximation.} because the dynamics has local maxima in the neighbourhood of which none of the curvature term may be neglected.
\item The second one is represented on the $B_1$ and $B_2$ graphs. It is a monotonic approach of the singularity.\\
\emph{It may be kasnerian as on $B_1$ graph}. It means that the functions $\alpha$, $\beta$ and $\gamma$ asymptotically behave as some linear decreasing functions of the $\tau$ time. This is due to the fact that, in the isotropy vicinity, the curvature terms and the potential become negligible with respect to the second derivative $\ddot\alpha$, $\ddot\beta$ and $\ddot\gamma$. The Bianchi type $IX$ model is then approximated by a Bianchi type $I$ one. This singularity approach is the same as the one of the Brans-Dicke theory in the Einstein frame when the scalar field energy density is positive($\omega>-3/2$, see appendix).\\
\emph{It may also be non Kasnerian} (but always monotonic) as on the figure $B_2$ where the singularity is reached for a finite value $\tau_0$ of the $\tau$ time. This singularity also exists in the proper time $t$ for a finite value $t=t_0$. Once again, the curvature terms may be neglected and the Bianchi type $IX$ model approximated by the Bianchi type $I$ one. However the potential is not negligible any more. It implies that $\alpha$, $\beta$ and $\gamma$ are not some linear functions of $\tau$. Thus the solution (which in general can not be calculated analytically) is not kasnerian. In particular this is the case with the power law potential $U=\phi^k$ with $k$ a negative integer and $\phi<0$ (see table \ref{tabDyn} for a numerical application).
\item The third one is shown on graphs $C_1$ and $C_2$. The oscillations end for a finite value $\tau_0$ of $\tau$ for which the Universe expands and the metric functions first derivatives diverge. The metric functions behaviours of these two figures are not physically equivalent. Indeed, in the proper time $t$, the graph $C_1$ transforms into graph $C_{11}$. We recover a singularity since the Ricci scalar diverges. However the graph $C_2$ casts into the graph $C_{21}$ where the singularity disappears. In this last case, the slopes of the three metric functions tend to the same value and the Universe isotropises.
\end{itemize}
Note that we have not detected any cigar or pancake singularity. It does not mean that they would not exist for other theories with some massive scalar fields. Moreover, whatever the functions $\omega$ and $U$, there is no isotropic point singularity. It is because, the difference between two first derivatives of $\alpha$, $\beta$ or $\gamma$ when the curvature terms are neglected is always a constant. It is why it does not exist for the $B_2$ figure in the $\tau$ time a similar behaviour as $C_{21}$ in the $t$ time. Some numerical examples of the seven metric functions behaviours illustrated on figure \ref{fig0} are given on table \ref{tabDyn}.
\begin{table}[!htbp]
\scriptsize
\begin{center}
\begin{tabular}{|c|c|c|c|c|c|c|c|c|c|c|c|}
\hline
Graph&$U$&$\alpha$&$\beta$&$\gamma$&$\phi$&$\dot\alpha$&$\dot\beta$&$\dot\gamma$&$\omega$&Param&Time\\
\hline
$A_1$&$\Lambda$&-0.51&-0.72&-0.91&0.23&-0.48&1.55&0.55&-1.75&$\Lambda=-1.0$&ET\\
$B_{1}$&$\Lambda$&-0.51&-0.72&-0.91&0.23&-0.48&1.55&3.55&2.1&$\Lambda=1.2$&ET\\
$B_{2}$&$\phi^k$&-0.51&-0.72&-0.91&-0.23&-0.48&1.55&3.55&-1.1&$k=-3.0$&ET\\
$C_{1(1)}$&$e^{k\phi}$&-0.51&-0.72&-0.91&0.23&-0.48&1.55&0.55&-2.5&$k=3.3$&ET\\
$C_{2(1)}$&$\Lambda$&-0.51&-0.72&-0.91&0.23&-0.48&1.55&3.55&1.1&$\Lambda=0.5$&LT\\
\hline
\end{tabular}
\caption{\label{tabDyn}\scriptsize{Initial conditions of some numerical applications illustrating the 7 metric functions behaviours of the figure \ref{fig0}. ET means that integration is to early times whereas LT means to late time.}}
\end{center}
\end{table}
\\Let us explain how we numerically solved the field equations to get the figures \ref{fig1}-\ref{fig3} where the above behaviours were observed. Each theory we consider is defined by a couple of control parameters appearing in the functions $\omega$, here taken as a constant $\omega_0$, and $U$. For a constant potential, it is $(\omega_0,\Lambda)$ and for a power or exponential law potential it is $(\omega_0,k)$. Each couple defines a plane. We integrated and determined the asymptotical behaviour of the metric functions for each of their points at early and late times. Numerical integrations of the field equations have been performed using a Runge-Kutta order 5 algorithm.\\
What about the initial conditions? If we look at the equation (\ref{eq5}), we see that it can be rewritten as:
$$
\dot{\phi}=\pm\sqrt{\left[\frac{3+2\omega}{\phi^2}\ddot\phi+e^{2(\alpha+\beta+\gamma)}U_\phi\right]\frac{\phi^{3}}{3+2\omega}}
$$
Hence we have used for each graph of figures \ref{fig1}-\ref{fig3} two sets of initial conditions $(\alpha_0,\dot\alpha_0,\beta_0,\dot\beta_0,\gamma_0,\dot\gamma_0,\phi_0)$ corresponding to the $\pm$ sign above. Moreover since each numerical integration uses different values of the control parameters, it starts with a different value of $\dot\phi_0$. To be sure that the properties observed on the figures \ref{fig1}-\ref{fig3} are stable, we have used several sets of initial conditions.\\
Now, let us explain the meaning of the graphs on the figures \ref{fig1}-\ref{fig3}. Basically each of them represents a plane with $\omega_0$ as ordinate and a second control parameter ($\Lambda$ or $k$) as abscissa. On each figure, the left(right) graphs represent the numerical integration results to early(late) times. The meaning of the textures covering the graphs is the following:
\begin{itemize}
\item The horizontal lines (first texture on figure \ref{carre}) will represent the points of the plane where a monotonic approach of the singularity arises. It may be kasnerian as on graph $B_1$ or not as on graph $B_2$ and the energy density is always positive. We have not distinguished these two cases because a numerical test will be too uncertain to discriminate them.
\item The vertical lines (second texture on figure \ref{carre}) will represent the points of the plane where an oscillatory behaviour of the metric functions occurs with a negative energy density. It correspond to graph $A_1$ on figure \ref{fig0}.
\item The tilted lines (third texture on figure \ref{carre}) will represent the points of the plane corresponding to an expanding Universe. It occurs for a finite value $\tau_0$ of the $\tau$ time, singular or not, with a positive energy density. It corresponds to graphs $C_1$ and $C_2$ on figure \ref{fig0}.
\item The points (fourth texture on figure \ref{carre}) will stand for the points of the plane where the asymptotical behaviour has not been clearly determined. Normally they should be horizontal line. However the monotonic approach of the singularity generally arises too slowly and an underflow occurs.
\item The white area will be the points of the plane for which the numerical tests have not been able to determine in an unquestionable manners, the asymptotical behaviour before an over or underflow. It is not possible to say anything about them.
\end{itemize}
\begin{figure}[h]
\centering
\includegraphics[width=10cm]{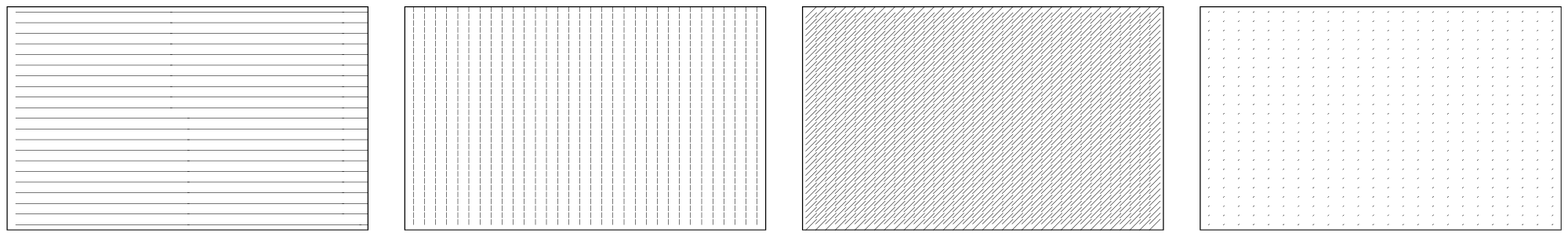}
\caption{\label{carre}\scriptsize{The four textures used in the figures \ref{fig1}-\ref{fig3} to represent the asymptotical behaviour of the Universe: contraction with a positive energy density, oscillatory behaviour with a negative energy density, expansion with a positive energy density, probably contraction with a positive energy density.}}
\end{figure}
In the next three subsections, we describe the results of the numerical integrations for each potential.
\subsection{Cosmological constant $U=2\Lambda$}\label{s21}
\begin{figure}[h]
\centering
\includegraphics[width=16cm]{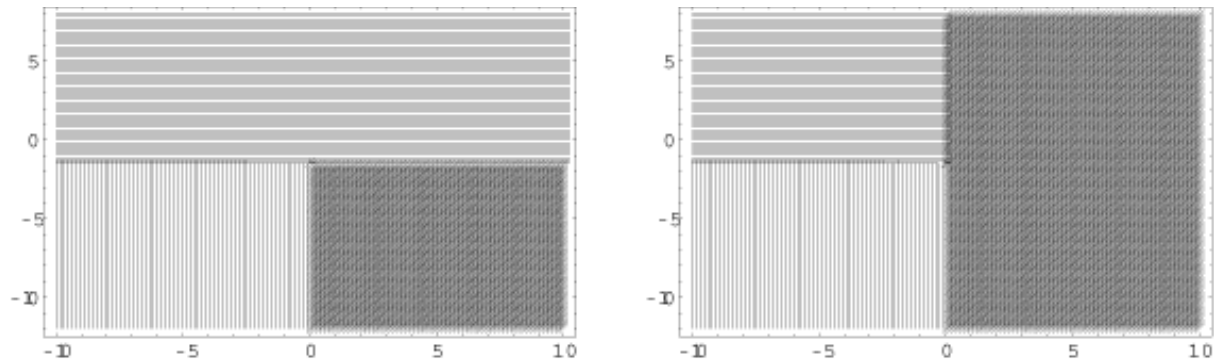}
\caption{\label{fig1}\scriptsize{Asymptotical behaviours of the Bianchi type $IX$ model when $\omega=\omega_0$ (in ordinate) and $U=2\Lambda$ ($\Lambda$ in abscissa). The left graphs correspond to early times and the right ones, to late times.}}
\end{figure}

Results of the numerical integrations are shown on figure \ref{fig1}. The abscissa stands for $\Lambda$ values and the ordinate for $\omega$ values. 
The metric functions asymptotical behaviours partly (but weakly) depend on a $\Lambda=\Lambda_0$ critical value related to the initial conditions. Summarising these behaviours, we get:
\begin{itemize}
\item When $\omega<-3/2$ and $\Lambda\leq 0$, the metric functions oscillate forever without tending to a singularity.
\item When $\omega<-3/2$ and $\Lambda>0$, the Universe is spatially open: the metric functions are always asymptotically expanding for a finite value $\tau_0$ of $\tau$ and have a minimum.
\item When $\omega>-3/2$ and $\Lambda\leq 0$, the Universe is spatially closed and asymptotically tends monotonically to a singularity, kasnerian or not, at early and late time. Metric functions have a maximum.
\item When $\omega>-3/2$ and $\Lambda>0$, it always exist a singularity at late(early) times. Then, at early(late) times, if $0<\Lambda<\Lambda_0$, the Universe also tends monotonically to a singularity and is thus spatially closed. If $\Lambda_0<\Lambda$, the Universe is expanding.
\end{itemize}
It is possible to partly explain this distribution of the asymptotical behaviours with respect to the control parameters $\omega_0$ and $\Lambda$. From the Klein-Gordon equation, it is easy to show that $\phi$ behaves as an exponential of $\tau$. Thus $\dot\phi^2\phi^{-2}$ is a positive constant. Consequently, when curvature terms are negligible in the constraint equation (\ref{eq4}), the term with the Brans-Dicke constant dominates. Then as for the Brans-Dicke theory, the metric oscillates when $\omega<-3/2$ because $\dot{\alpha} \dot{\beta} +\dot{\alpha} \dot{\gamma} +\dot{\beta} \dot{\gamma}<0$. On the contrary when $\omega>-3/2$, the three first derivatives may have the same sign and thus the metric functions approach to singularity is monotonic. When the Universe expands, the numerical simulations show that the dominant terms in the equation (\ref{eq4}) are the first derivatives of the metric functions and the potential. Consequently, when the three metric functions expand, all the first derivatives have the same sign and $\Lambda$ must be positive. This is what occurs on figure \ref{fig1}\\
The asymptotical behaviour of the metric functions is thus weakly related to the initial conditions but to the signs of $3+2\omega$ and $\Lambda$. This is due to the fact that the potential does not depend on the scalar field. In what follows, we are going to consider some potentials depending on $\phi$. Then we will check that the metric functions asymptotical behaviour strongly depends on the initial conditions although some properties, such as oscillations, stay partly ruled by the control parameters.
\subsection{Power law of the scalar field $U=\phi^k$}\label{s22}
\begin{figure}[h]
\centering
\includegraphics[width=16cm]{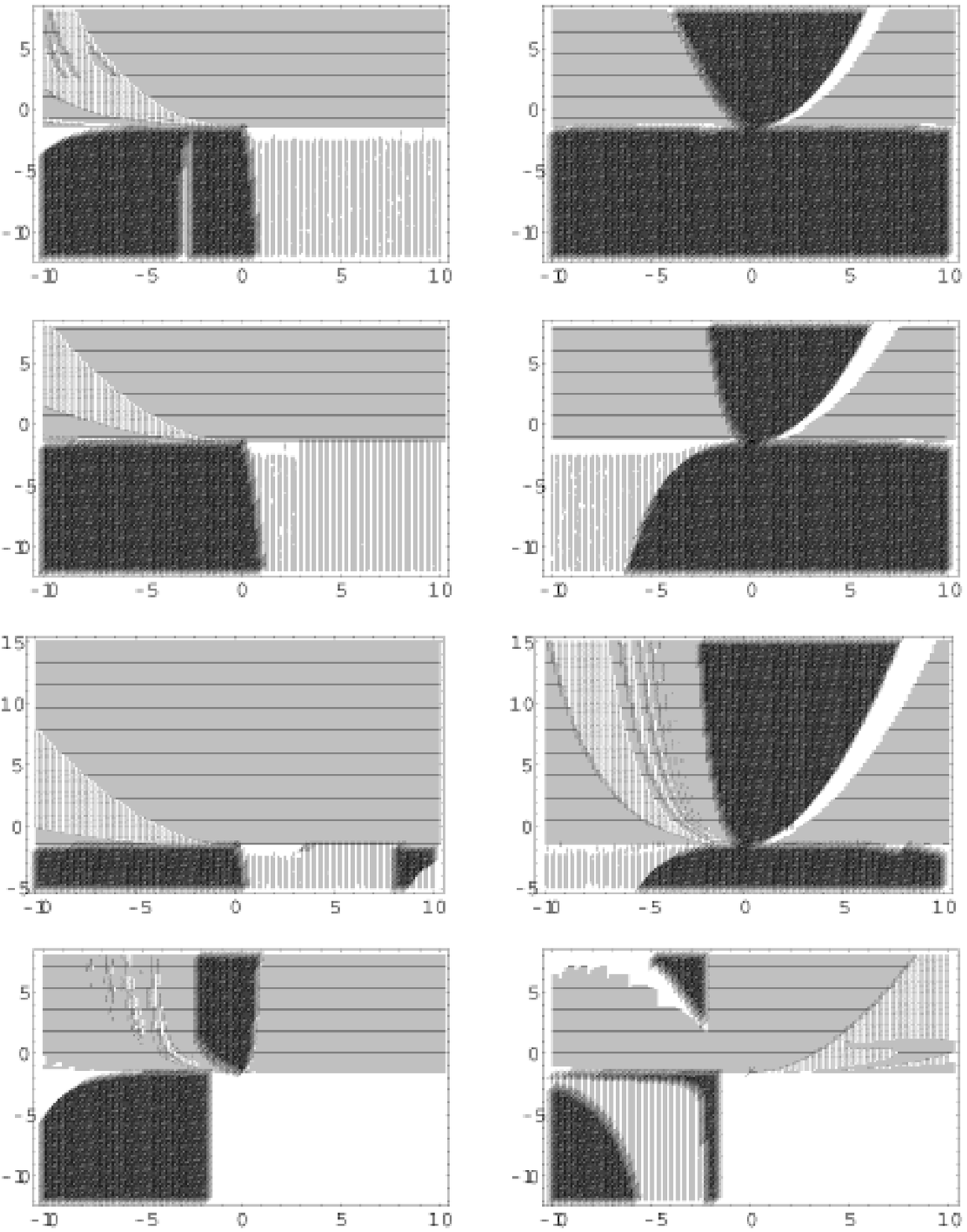}
\caption{\label{fig2}\scriptsize{Asymptotical behaviours of the Bianchi type $IX$ model with a theory defined by $\omega=\omega_0$ (in ordinate) and $U=\phi^k$ ($k$ in abscissa) when $k$ is real. The left graphs represent the early times and the right graphs, the late times.}}
\end{figure}

Numerical results are shown on figure \ref{fig2} with $k$ in abscissa and $\omega$ in ordinate. We begin to consider non integer values of $k$. Consequently, we choose a positive initial value for the scalar field. This guarantees that the potential is real.\\
One more time, when $\omega>-3/2$, the metric functions never oscillates. They monotonically decrease to a singularity or increase at late or early times depending on the initial conditions. 
Some solutions have both a Big-Bang and a Big-Crunch and are thus spatially closed. Others begin(end) with a singularity and end(begin) by an expansion. In any case, the singularity approach is monotonic. Solutions do not exist that are spatially open at both early and late times. The set of parameters couples $(\omega,k)$ corresponding to an expanding Universe has a hyperbolic form and is centred on the ordinate axis. It is explained in part by calculating the necessary condition such that the Universe becomes isotropic and expanding. Using \cite{Fay04A}, it writes $\omega<\frac{1}{2}(k^2-3)$ and is valid when $3+2\omega$ and $U$ are positive. It is the equation of an hyperbole in the $(\omega,k)$ plane with a minimum in $(\omega,k)=(-3/2,0)$ in agreement with the figure \ref{fig2}. Note that all the points of the hyperbole do not necessarily correspond to an isotropic Universe as explained in section \ref{s2} with $C_1$ and $C_2$ graphs.\\
When $\omega<-3/2$, the metric functions oscillate or increase at early or late times, depending on the initial conditions. Some solutions have a minimal 3-volume. There is no solution which asymptotically oscillates at early and late times contrary to what happens in presence of a cosmological constant or for a massless Brans-Dicke theory. It shows that the positive potential ends to dominate the negative kinetic term of the scalar field. It thus destroys the oscillations and leads to a monotonic behaviour.\\
\begin{figure}[h]
\centering
\includegraphics[width=16cm]{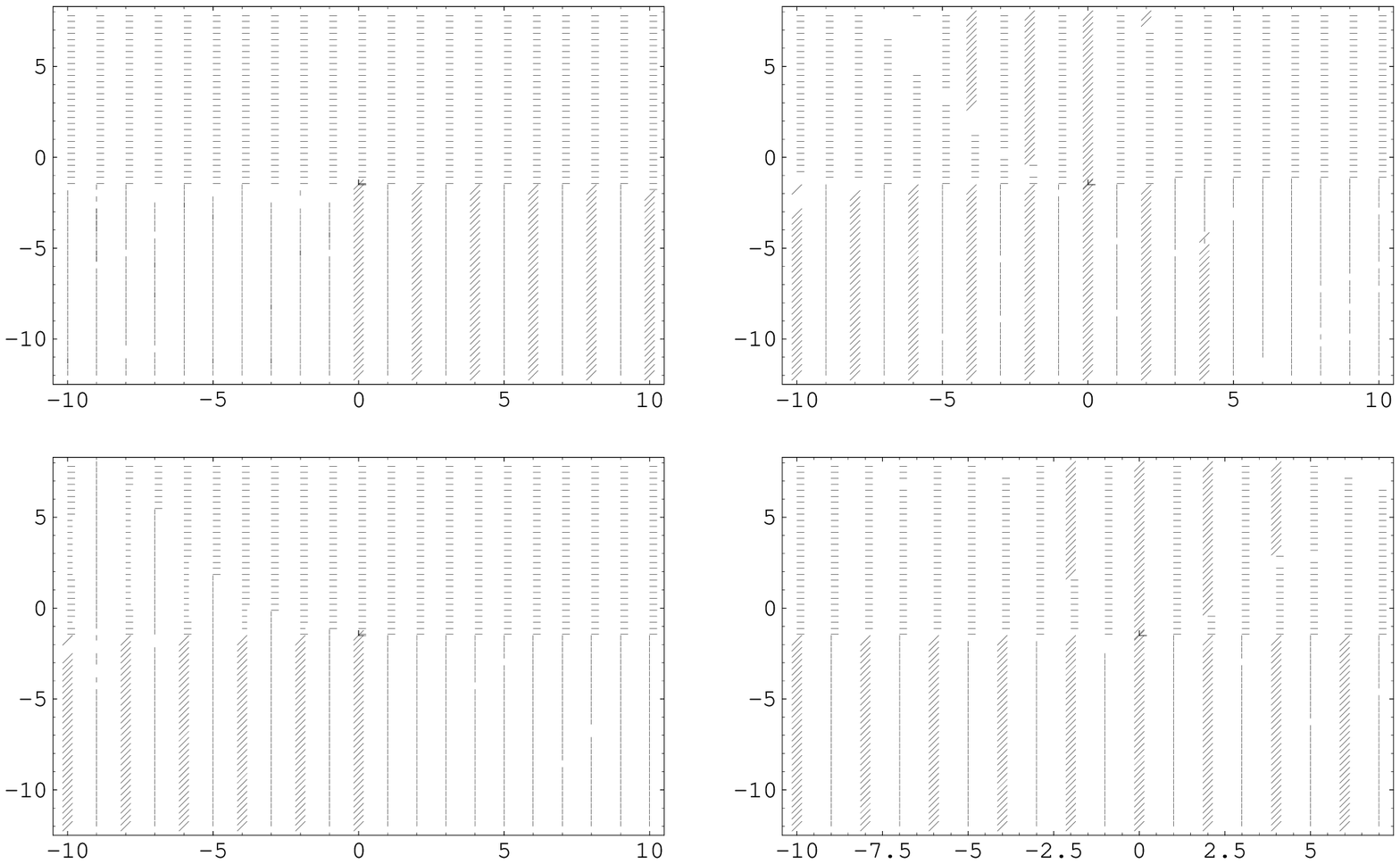}
\caption{\label{fig2a}\scriptsize{Asymptotical behaviours of the Bianchi type $IX$ model with a theory defined by $\omega=\omega_0$ (in ordinate) and $U=\phi^k$ ($k$ in abscissa) when $k$ is an integer. The left graphs represent the early times and the right graphs, the late times.}}
\end{figure}
What happens when $k$ is an integer? The asymptotical behaviour of the metric functions then partly depends on the $k$ parity as shown on figure \ref{fig2a}. In order to find some results different from the previous one, we choose a negative initial value for the scalar field. The two upper graphs are such that the root of $\dot\phi^2$ is positive, and the two other ones with the negative root.\\
When $\omega_0>-3/2$, again, there is no oscillation. The Universe is expanding or monotonically approaches a singularity. This is not an obvious result, because the negative potential could dominate the scalar field kinetic terms thus implying some oscillations. However, it does not occur. We see that an expanding Universe only appears for even values of $k$ corresponding to a positive potential as above when we considered real values of this parameter. As opposed to this case, the monotonic approach (kasnerian or not) of the singularity does not depend on $k$ being even or not.\\
When $\omega_0<-3/2$, the Universe is either expanding or oscillating. It never contracts to a singularity. Once again, expanding Universe only appears for even value of $k$, i.e. positive potential, whereas oscillating behaviour of the metric functions does not depend on $k$ parity. However, stable oscillating solutions at both early and late times only appear for odd values of $k$ whereas some solutions having a minimum 3-volume only arise for even values of this parameter.\\
This  application shows in an obvious way the dependence of the Universe asymptotical behaviour on the sign of the scalar field and thus of the potential.
\subsection{Exponential law of the scalar field $U=e^{k\phi}$}\label{s23}
\begin{figure}[h]
\centering
\includegraphics[width=16cm]{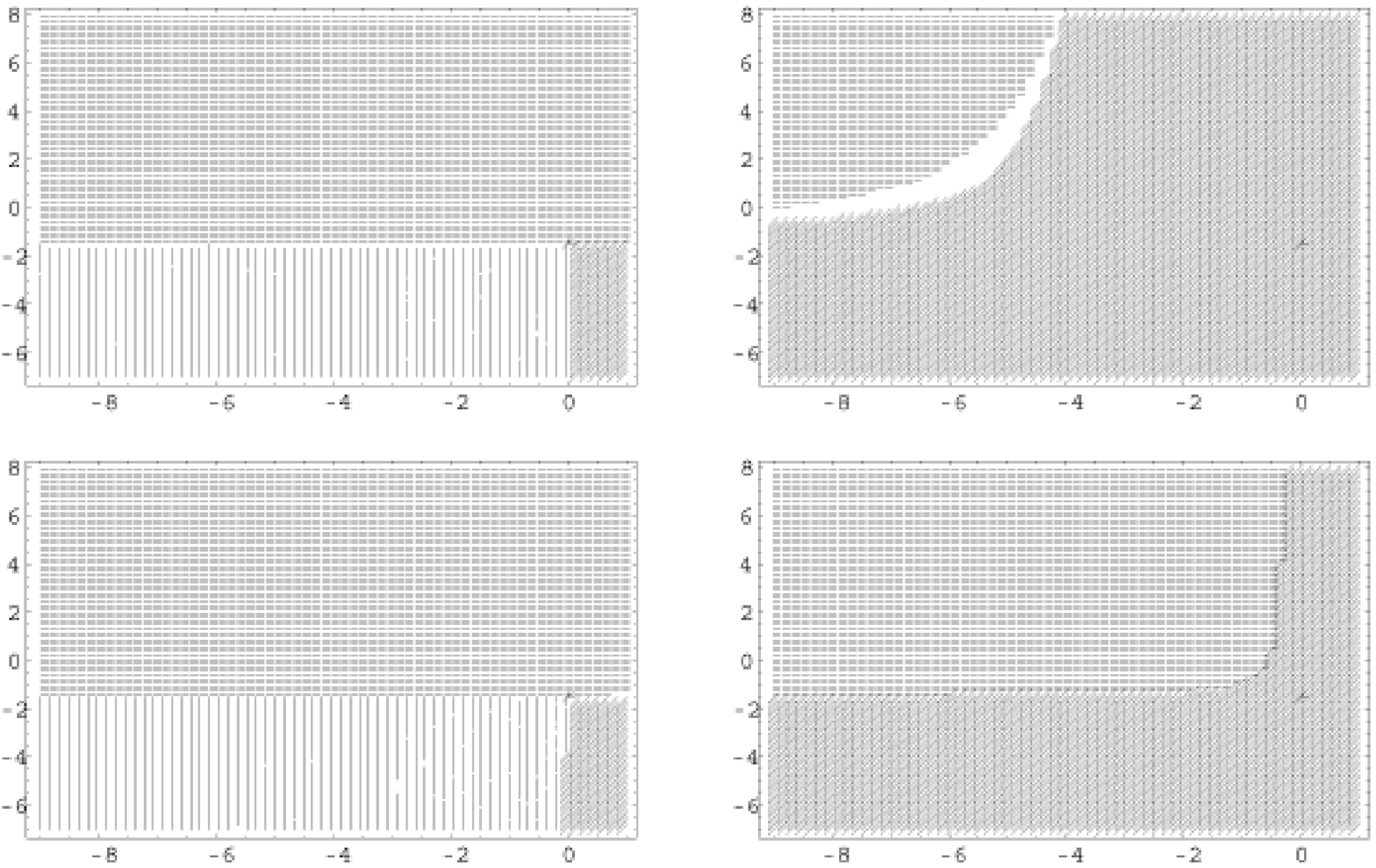}
\caption{\label{fig3}\scriptsize{Asymptotical behaviours of the Bianchi type $IX$ model with a theory defined by $\omega=\omega_0$ (in ordinate) and $U=e^{k\phi}$ ($k$ in abscissa). The left graphs represent the early times and the right graphs, the late times.}}
\end{figure}

The results of the numerical integrations are shown on figure \ref{fig3} with $k$ in abscissa and $\omega_0$ in ordinate. 
These figures have been computed assuming $\dot\phi_0<0$. Apparently the metric functions oscillations only occur for $k<0$ at early times. But if we choose a positive $\dot\phi_0$, it also arises for $k>0$ and again at early times. We have not observed such behaviours for late time. Only two properties seem independent from the initial conditions. The fist one is a singularity at early times when $\omega>-3/2$. It is reached monotonically by the metric functions. The second one is the Universe expansion at late times when $\omega>-3/2$ and $k>0$.\\
For the other sets of $\omega$ and $k$ values, the asymptotical state depends on the initial conditions.\\
When $\omega>-3/2$ and $k<0$, it exists some solutions having a Big-Bang and Big-Crunch whose approach is monotonic. No oscillatory behaviour has been detected. Some solutions with a Big-Bang at early times and which expands at late time have been detected but never the opposite. There is no solution which is expanding both at early and late times and thus has a minimum 3-volume.\\
When $\omega<-3/2$ and $k<0$, the early and late times are such that the Universe is expanding or oscillating. Whatever $k$, it exists some solutions having a minimal 3-volume. It does not exist any solution oscillating at both early and at late times, indicating that this state is unstable.\\
\\
To conclude, this section shows that, contrary to what happens with the Brans-Dicke theory in the Einstein frame, the asymptotical behaviour of the metric functions also depends on the initial conditions and not only on the control parameters. However, it is always possible to find some properties which are independent from the initial conditions, in particular with a cosmological constant. As instance, forever oscillating Universe or Universe having a minimum volume.
\section{Positive energy density and metric oscillations}\label{s3}
\begin{figure}[h]
\centering
\includegraphics[width=6cm]{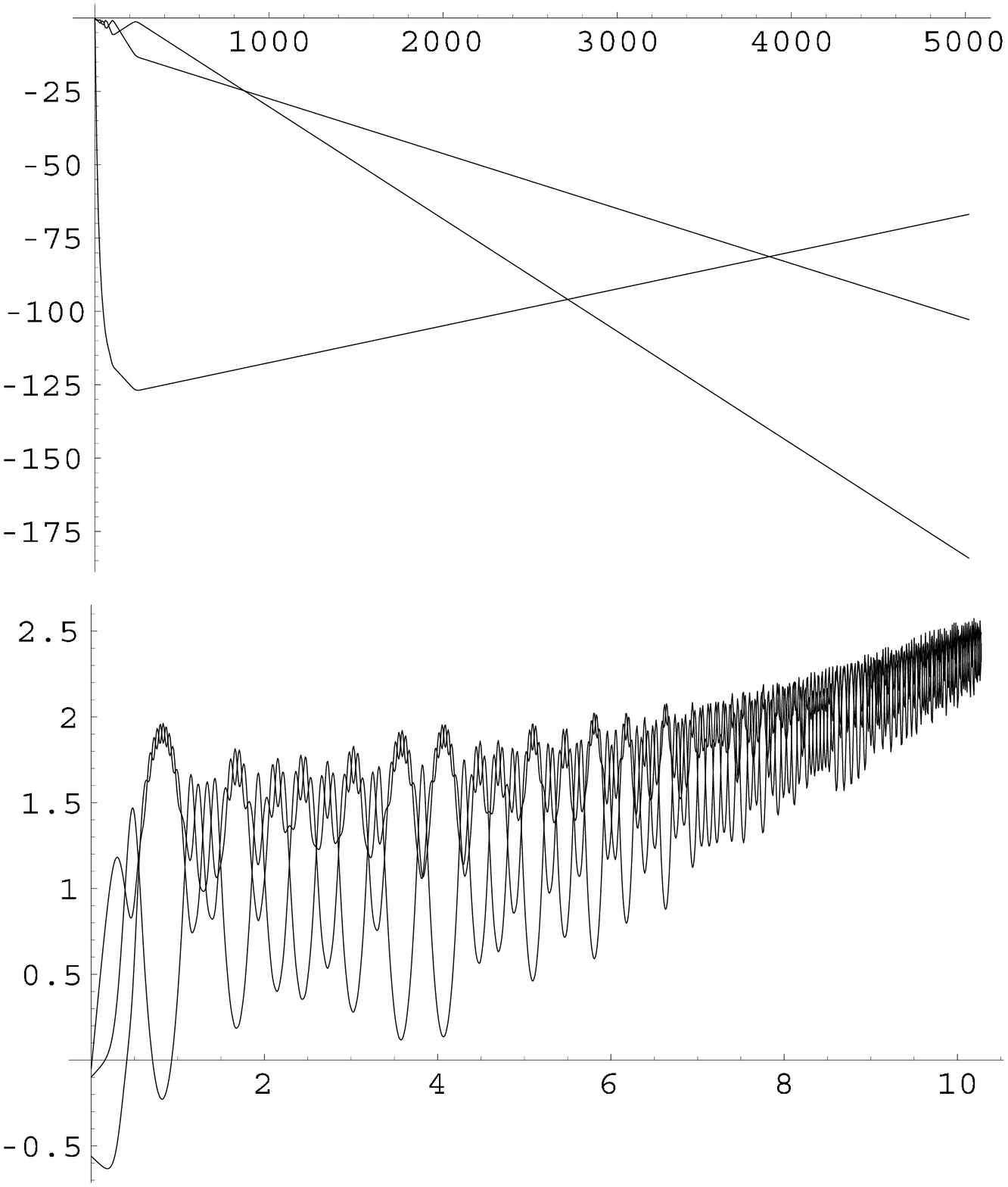}
\caption{\label{fig4}\scriptsize{Asymptotical behaviours of the Bianchi type $IX$ model for the theory defined by $\rho_\phi V^2=k_ 1 e^\tau>0$. The initial conditions are $\alpha=-0.56$, $\beta=-0.098$, $\gamma=-0.048$, $\dot\alpha=0.48$, $\dot\beta=0.55$, $\dot\gamma=5.23$ and $k_1=0.35$. The first graph is got by integrating to early times and the second one to late times.}}
\end{figure}

One of the interest of the Bianchi type $IX$ model is its singularity oscillatory approach\cite{Mis69}. It is able to isotropise the Universe or iron out inhomogeneities. Such a singularity exists in General Relativity in presence of a perfect fluid of dust or radiation whose effects are then negligible. In presence of a massless scalar field, things are different. When the energy density is positive or equivalently the weak energy condition is respected ($\omega>-3/2$), the singularity approach is kasnerian and monotonic. Otherwise($\omega<-3/2$), the metric functions oscillate as on graph $A_1$ but the Universe does not go to a singularity (see appendix for heuristic arguments with the BKL approximation or the next sections). Apparently, the situation could be the same with a massive scalar field: the three basic theories we worked out in the previous section always show a monotonous (kasnerian or not) approach of the singularity and never an oscillatory one. If it was true, it would mean that a scalar field, massless or massive, kills the oscillatory approach of the Bianchi type $IX$ singularity. Here, we show using a toy model that this assumption is wrong: \emph{1 - singularity oscillatory approach may exist in presence of a massive scalar field}. Moreover, \emph{2 - it does not require a negative scalar field energy density but the non respect of the weak energy condition could be necessary}. This difference between "a negative scalar field energy density" and "the non respect of the weak energy condition" only exists for a massive scalar field. For a massless one, there is no difference. The potential thus plays an important role.\\
To show these two important results, first we write the density and pressure of the scalar field as:
$$
\rho_\phi=\frac{1}{2}V^{-2}(\frac{3}{2}+\omega)\frac{\dot\phi^2}{\phi^2}+\frac{1}{2}U
$$
$$
p_\phi=\frac{1}{2}V^{-2}(\frac{3}{2}+\omega)\frac{\dot\phi^2}{\phi^2}-\frac{1}{2}U
$$
and we define the following variables:
$$
K_1=V^2\rho_\phi
$$
$$
K_2=\frac{\dot V}{V}(\rho_\phi-p_\phi)\rho_\phi^{-1}
$$
$V$ being the comoving 3-volume $e^{\alpha+\beta+\gamma}$. $K_1$ has no particular meaning but is a positive quantity. When $p_\phi/\rho_\phi$ tends to a constant, $K_2$ is proportional to the variation of the comoving 3-volume in the proper time. Using $K_1$ and $K_2$, the field equations write:
\begin{eqnarray*}
\ddot{\alpha}&=&\frac{1}{2}\left[(e^{2\beta}-e^{2\gamma})^2-e^{4\alpha}\right]+\frac{1}{2}\frac{V}{\dot V}K_1K_2\\
\ddot{\beta}&=&\frac{1}{2}\left[(e^{2\alpha}-e^{2\gamma})^2-e^{4\beta}\right]+\frac{1}{2}\frac{V}{\dot V}K_1K_2\\
\ddot{\gamma}&=&\frac{1}{2}\left[(e^{2\alpha}-e^{2\beta})^2-e^{4\gamma}\right]+\frac{1}{2}\frac{V}{\dot V}K_1K_2\\
\dot{\alpha} \dot{\beta} +\dot{\alpha} \dot{\gamma} +\dot{\beta} \dot{\gamma} &=&\frac{1}{4}\left[e^{4\alpha}+e^{4\beta}+e^{4\gamma}-2(e^{2(\alpha+\beta)}+e^{2(\alpha+\gamma)}+e^{2(\beta+\gamma)})\right]+\\
K_1\\
\dot K_1=K_2K_1\\
\end{eqnarray*}
We then define our toy model as $K_1=k_1e^\tau>0$ where $k_1$ is a positive constant. The scalar field energy density is thus positive and corresponds to $K_2=1$, i.e. the comoving 3-volume asymptotically behaves as $t$. We numerically integrate the field equations to get the two graphs of the figure \ref{fig4}. 
The first one is obtained by integrating to early times. Then, the scalar field energy density tends to vanish and the metric approach a singularity by oscillating. The second one is got by integrating to late times. $\rho_\phi$ and the comoving 3-volume increase as the metric functions oscillate faster. It allows to the metric functions derivatives to counterbalance the increasing of $\rho_\phi V^2$ which stays negligible in the field equations. It is not possible to recover analytically the forms of $\omega$ and $U$. However, we can check numerically that $\omega$ is not a constant and is always smaller than $-3/2$. It means that it is the potential which keeps the energy density positive all along the integration and that the weak energy condition is violated.\\
Hence our toy model allows an oscillatory singularity approach with a massive scalar field having a positive energy density but violating the weak energy condition.
\section{Hamiltonian analysis}\label{s4} 
In this section we would like to use a simple method to foretell the presence of chaos or not in the various systems that we studied above, in particular by making predictions in the $(\omega _{0},k)$ plane.
\subsection{Chaos energy threshold}
We shall use an approach of chaos in term of energy threshold. It has been shown by several methods (numerically and analytically, see for example a recent study by \cite{LehMen03}) that such a threshold of chaos exists in the Bianchi type $IX$ vacuum Universe case but also in presence of massless scalar fields.
\newline
\newline
In the vacuum case, the chaos threshold is reached when the Hamiltonian density
$$
h(\alpha ,\beta ,\gamma)=e_{c}+e_{p}
$$
with
$$
e_{c}=\dot{\alpha}\dot{\beta}+\dot{\alpha}\dot{\gamma}+\dot{\beta}\dot{\gamma}
$$
$$
e_{p}=\frac{-1}{4}\left[ e^{4\alpha }+e^{4\beta}+e^{4\gamma}-2(e^{2(\alpha +\beta )}+e^{2(\alpha +\gamma )}+e^{2(\beta+\gamma)})\right]
$$
is vanishing. This threshold is precisely realized since one has exactly $h=0$. 
\newline
In presence of ordinary matter, the curvature terms dominates the matter ones that behave as $\rho V^{2}$. Asymptotically there is a singularity implying a vanishing volume V. Thus we get again in this limit $h=0$ with a chaotic approach to the singularity.
\newline
In the case of stiff matter, the matter component adds a contribution $\rho V^{2}$ that scales like $\rho_{0}V^{(2-\gamma )}$ with $\gamma=2$. Thus the constraint equation at any time is now $h-\rho_0=0$. For $\rho_0>0$ we find that the chaos is stopped since $h>0$ (we stay below the threshold). But for an exotic matter such that $\rho_0<0$ we get $h<0$ and we enter in a chaotic regime.
\newline
The same conclusion applies in the case of a massless scalar field $\phi$, equivalent to a stiff matter. It adds a kinetic contribution to the Hamiltonian density such that the constraint reads now h=$\frac{1}{2}\frac{\dot{\phi}^{2}}{\phi ^{2}}(\omega +\frac{3}{2})$. Hence for $\omega>-\frac{3}{2}$ there is no chaos but there is for $\omega <-\frac{3}{2}$. Other cases have been studied including both ordinary matter and a scalar field, leading to the same conclusion\cite{LehMen03}.
\newline
\newline
The problem here is to predict what is going on with a massive scalar field, but without ordinary matter. The corresponding Hamiltonian density for the scalar field reads :
$$
H(\phi)=E_{c}(\phi )+E_{p}(\phi)
$$
with
$$
E_{c}(\phi)=-\frac{1}{2}\frac{\dot{\phi}^{2}}{\phi ^{2}}(\omega _{0}+\frac{3}{2})
$$
$$
E_{p}(\phi)=-\frac{1}{2}UV^{2}=-\frac{1}{2}e^{2(\alpha +\beta +\gamma )}U
$$
and the constraint for the total system is thus
\begin{equation}  \label{consEq}
h+H(\phi )=0
\end{equation}
\noindent 
From equation (\ref{consEq}) chaotic situations can be predicted in several cases since here the chaos condition reads h=-H($\phi)<0$. Simple predictions will be made in presence or in absence of volume singularity (see next section) and are summarized in the section entitled general predictions.
\subsection{Volume singularity analysis}
An useful item is to compute the volume time evolution. For that, we introduce the time variable $\xi $ defined by 
$$
d\xi =\frac{dt}{V\phi }=\frac{e^{-(\alpha +\beta +\gamma )}}{\phi }dt
\label{tauprime}
$$
Adding the three differential equations for $\alpha $, $\beta $ and $\gamma $, and using the constraint equation (\ref{eq4}), one generally gets the following result for $u=\ln (V)=\alpha +\beta +\gamma $:
\begin{eqnarray}
u_{\xi\xi}&=&-2e_{p}+3Ue^{2u}/2\label{vol}\\
u_{\xi\xi}-u_\xi^2 &=&-\left[ \alpha_\xi^2 +\beta_\xi^2+\gamma_\xi^2\right] +2H(\phi )+3Ue^{2u}/2=... \\
... &=&-\left[ \alpha_\xi^2+\beta_\xi^2+\gamma_\xi^2\right] +\frac{1}{2}e^{2u}U(\phi )-\frac{\dot{\phi}^{2}}{\phi ^{2}}(\omega_{0}+\frac{3}{2})\label{volp}
\end{eqnarray}
where a subscript $\xi$ means a derivative with respect to $\xi$. This last equation can help us to predict cases that are not covered by the direct energy analysis.\\
In the vacuum Universe, equation (\ref{volp}) reduces to 
\begin{equation}
\ddot u=-2e_{p}=-\left[ \dot\alpha^2+\dot\beta^2+\dot\gamma^2\right]\label{volvide1}
\end{equation}
and we have besides the relation:
\begin{equation}
\ddot u-\dot u^2=V^2d_{tt}^{2}(LnV)  \label{volvide2}
\end{equation}
Thus in the vacuum case if the volume starts to decrease, it will decrease forever according to equations (\ref{volvide1}-\ref{volvide2}), indicating a singularity.\\
\\
In presence of the massive scalar field we are back to the full equation (\ref{volp}) and we can state the following:\\
\\
If $U<0$ and $\omega_{0}+\frac{3}{2}>0$, and if the volume starts to decrease, it will decrease again forever. The same conclusion applies about its possible increase but also for $u_\xi\phi_\xi$ going to zero with time since we have the relation:
\begin{equation}
u_{\xi\xi}-u_\xi^2=V^{2}\phi ^{2}d_{tt}^{2}(LnV)+u_\xi\phi_\xi\label{volrel}
\end{equation}
Thus now the quantity $u_{\xi\xi}-u_\xi^2$ is not directly connected to the sign of the second derivative in time $t$ of $Ln(V)$ due to the coupling term in $u_\xi\phi_\xi$ and to the definition of $\xi$. Hence the conclusions valid in vacuum should be adapted including the relative sign of $u_\xi$ and $\phi_\xi$.
\begin{itemize}
\item For $u_\xi\phi_\xi>0$ the conclusion remains the same than above about a possible volume singularity.
\item For $u_\xi\phi_\xi<0$ the full dynamics should be solved in order to compare the relative magnitude of the negative terms in the right hand side (rhs) of equation (\ref{vol}) and the positive term $-u_\xi\phi_\xi$.
\end{itemize}
However there is not always a volume singularity.
\subsection{General predictions}
Our numerical studies have shown that there is not always a singularity: in particular the Universe volume can be oscillating without going to zero. It then exhibites some local minima corresponding to some bounces\cite{TopUst99}. Following \cite{LehMen03}, the chaos condition reads now $h=-H(\phi)=\rho _{\phi }V^{2}<0$ while the weak energy condition is $p_{\phi}+\rho _{\phi }=-2E_{c}(\phi )/V^{2}$. Hence it comes:
\begin{itemize}
\item When the volume singularity exists i.e $V\rightarrow 0$ for infinite $\tau$ time, for
$U$ bounded in time $UV^{2}\rightarrow 0$ and $-H\rightarrow -E_{c}(\phi )$ (unless the factor $\frac{\dot\phi}{\phi}$ in $E_c$ vanishes faster than $UV^2$).

Thus there is chaos for $\omega_{0}<-\frac{3}{2}$ whatever the sign of $U$, the weak energy condition  being violated and the energy density being negative. There is no chaos for $\omega_{0}>-\frac{3}{2},$ the weak energy condition being respected and $\rho_{\phi }>0.$

For $UV^{2}$ not bounded with time no direct conclusion can be drawn.

\item When there is volume singularity or not:

a) for $U>0$ and $\omega_{0}>-\frac{3}{2}$ there is always no chaos, the weak energy condition is respected and $\rho_{\phi }>0$

b) for $U<0$ and $\omega _{0}<-\frac{3}{2}$ there is chaos with violation of the weak energy condition and $\rho _{\phi }<0$

c) for $U(\omega _{0}+\frac{3}{2})<0$ there is no direct conclusion. The weak energy condition is violated if $\omega<-3/2$, whatever the sign of the potential whereas the sign of $\rho_{\phi }$ is undetermined.
\item  Without volume singularity:

In this case we have a necessary condition that comes from $u_{\xi\xi}-u_\xi^2-u_\xi\phi_\xi>0$ which can be split into $2E_{c}(\phi )-E_{p}(\phi )-u_\xi\phi_\xi>0$. But nothing else can be said.
\end{itemize}
\subsection{Applications}
Let us make applications for the three specified potentials.
\subsubsection{case 1 : $U=2\Lambda $}
One concludes from above and also from the constraint equation (\ref{consEq}) that for $\Lambda >0$ and $\omega _{0}>-\frac{3}{2}$ there should be no chaos in the approach to the singularity. While for the case $\Lambda <0$ and $\omega _{0}<-\frac{3}{2}$ one should observe chaos. This is nicely
confirmed by the numerical simulations(see figure \ref{fig1}).\\
As explained in \ref{s21}, $\phi\propto e^{\phi_0\phi}$, with $\phi_0$ a constant. Hence $\dot\phi\phi^{-1}=\phi_0$ and, when there is a singularity (the volume is going there to zero), the scalar field kinetic energy will always dominate the potential energy on the long time. Thus, in this case, the conclusion for the chaos remains the same with respect to the value of $\omega _{0}$ with respect to $-\frac{3}{2}$, whatever the sign of $\Lambda$.
\newline
Undetermined situations about chaos occur when there is no singularity (the volume may be in expansion) and when $\Lambda(3+2\omega<0)$. Note that in some cases each variable $\alpha _{i}$ may be chaotic while their sum (related to the volume logarithm) could be regular.
\subsubsection{case 2: $U=\phi^{k}$}
We first consider the case of a real $k$. Again from the Hamiltonian constraint equation (\ref{consEq}), one concludes that for $\omega _{0}>-\frac{3}{2}$ one should have no chaos, since the fact to take $k$ real imposes $\phi>0$. For $\omega _{0}<-\frac{3}{2}$ we have no net conclusion
since the signs are opposite for $E_{c}(\phi )$ and $E_{p}(\phi )$.
\newline
Now we consider integer values for $k$. For $k$ even integer, we expect the same behaviour than for $k$ real with $\phi>0$. But for $k$ odd, we can take $\phi <0$ and thus a different behaviour of the system is expected : this is well confirmed by the numerical simulations (see figure \ref{fig2a}),
showing that the notion of energy threshold still holds here.%
\subsubsection{case 3: $U=e^{k\phi }$}
Again we are led to the same conclusion than above: there is no chaos in the case $\omega_{0}>-\frac{3}{2}$ whatever the sign of $k$ since the potential is positive. However no clear conclusion can be further drawn when $\omega_{0}<-\frac{3}{2}$ since $3+2\omega_0$ and $U$ have opposite signs.

Nothing can be found in the other cases.
\newline
When there is a singularity with the volume going to zero, we try to estimate the value of $e^{2u}U$ from equations (\ref{vol}-\ref{volp}). For instance, for $k<0$ and $\phi $ being an increasing function of time, the potential term will go to zero and the kinetic energy will dominate at late
time (idem for the case $k>0$ if $\phi $ can decrease with time). Then we get the same conclusion as stated above (for a  bounded potential energy with time).
\section{Discussion}\label{s6}
In this work we studied the dynamical behaviour of the Bianchi type $IX$ model in General Relativity with a massive scalar field. We consider three forms of potential. A first question was to know if, as for the massless Brans-Dicke theory in the Einstein frame, this behaviour is ruled only by some control parameters or if the initial conditions also play a role. A second question was to know if a scalar field always kills the singularity oscillatory approach.\\
To answer to the first question, we began to identify three different forms of asymptotical behaviours which may be divided in subcases:
\begin{enumerate}
\item The Universe expands in all the directions with $\rho_\phi>0$. This expansion may be singular, i.e. infinite and anisotropic for a finite proper time, or lead the forever expanding Universe to isotropy.
\item The Universe contracts monotonically and the curvature terms are negligible. Then the solution of the field equations may be kasnerian if the potential of the scalar field is also negligible. In this case the metric functions behave like power laws of the proper time. It may also be non kasnerian if the potential is not negligible. In any case, one has $\rho_\phi>0$ and $3+2\omega>0$.
\item The Universe oscillates without going to a singularity with $\rho_\phi<0$ and $3+2\omega<0$.
\end{enumerate}
Contrary to the Brans-Dicke theory in the Einstein frame with a massless scalar field, the metric functions behaviour does not depend only on the control parameters but also on the initial conditions. However, for each theories, there are some general properties which are ruled only by the control parameters. In particular when one considers a theory with a cosmological constant: its asymptotical behaviour seems only related to the initial conditions via a special value $\Lambda_0>0$ of this constant when $3+2\omega>0$. Otherwise, it only depends on the signs of $\Lambda$ and $3+2\omega$. This weak dependence with respect to the initial conditions occurs because $\omega$ and $\Lambda$ are some constants. This theory is thus probably the less initial conditions depending one when we consider a massive scalar field.\\
For the other forms of potentials which depend on $\phi$, the distribution of the metric functions asymptotical behaviour in the control parameters plane shares some common properties with the cosmological constant one:
\begin{itemize}
\item Oscillatory behaviours only occur when $\omega<-3/2$. Then the weak energy condition is violated. The test of chaos resting upon the Hamiltonian formalism developed in \cite{LehMen03} and used in the previous section goes in this way.
\item Spatially open Universes at late and early times only occur when $\omega<-3/2$ and $U>0$.
\item Closed Universes at late or early times only occur when $\omega>-3/2$.
\item Continuously oscillating Universes only occur for $\omega<-3/2$ and $U<0$.
\end{itemize}
However, on the contrary to the cosmological constant case, the repartition of the expanding or oscillating Universes in the control parameters planes restricted by the above conditions strongly depends on the initial conditions. This is due to the dependence of the potential with respect to the scalar field.\\
To our knowledge, this is the first time that figures such as \ref{fig1}-\ref{fig3} are obtained. They show a new way to study numerically the asymptotical behaviour of a Bianchi type $IX$ model in General Relativity with a massive scalar field. It is a new possibility to determine its properties in the same spirit as the Brans-Dicke theory in the Einstein frame with respect to the Brans-Dicke coupling constant $\omega$. A similar numerical study was performed in \cite{TopUst99} with the potential $U=m^2\phi^2$. The sets of initial conditions leading to bouncing solutions was analysed using a two dimensions slices through the initial conditions space.\\
Another common property for the theories of these papers is that an oscillating behaviour always needs a negative scalar field energy density, does not respect the weak energy conditions and never lead to a singularity. An important question is thus to know if an oscillatory singularity approach is always possible despite the presence of a massive scalar field. To answer, we rewrote the field equations and looked for a particular theory whose scalar field energy density becomes sufficiently small to be dynamically negligible. It is what happens in General Relativity with a perfect fluid for which the singularity approach is oscillatory. We found such a theory and, thanks to the potential which allows distinguishing between a violation of the weak energy condition and the presence of a negative scalar field energy density, we showed that a singularity oscillatory approach was possible even with a massive scalar field having a positive energy density but always violating the weak energy condition.\\
A next step, would be to prove that it also arises when $\rho_\phi$ diverges, which is the behaviour we are waiting for any kind of matter in the neighbourhood of a singularity (if we do not take into account any quantum effect). It should be possible since, in the field equations, $\rho_\phi$ is multiplied by the comoving 3-volume which should then disappear. It would also be interesting to test nonminimally coupled scalar field or other forms of potential such as the supergravity one. Moreover some Painlevé tests could be used to analyse the singularity from the integrability point of view \cite{DemSch96,Sch97}.
\section*{Acknowledgments}
We thank very much Doctor Christian Scheen for useful discussions on Bianchi type $IX$ model and scalar-tensor theories and for its reading of the manuscript. Many thanks to a good friend for having computed the figures of this paper.
\appendix
\section{The Belinskii, Khalatnikov and Lifshitz (BKL) approximation}
The BKL approximation\cite{BelKhaLif70} allows understanding easily with heuristic arguments why the metric oscillates or not. Explanations.
\subsection{In General Relativity}
The BKL approximation allows a qualitative description of the metric functions behaviour for the Bianchi type $IX$ model in the vicinity of the singularity.\\
One assumes that initially the metric functions are very small. In this case, curvature terms appearing in the field equations are negligible. These equations can then be approximated by those of the Bianchi type $I$ model whose solution is kasnerian: the metric functions behave as powers of the proper time $t$ such as their exponents $p_i$ have the properties $\sum_{i=1}^3p_i=\sum_{i=1}^3p_i^2=1$. These equalities imply that two of the metric functions decrease while the third one grows. It follows that after a finite time, one of the curvature terms, say $e^{\alpha}$, is not negligible any more. Then, the BKL approximation is invalid and the field equations approximated by that of the Bianchi type $II$ model. However, asymptotically this model tends to a Bianchi type $I$ one and thus, the BKL approximation becomes valid again. Repeating this process, the Bianchi type $IX$ model is thus successively approximated by some Bianchi type $I$ and $II$ models and the metric functions oscillate in the vicinity of the singularity. The interval of time between two oscillations is called a Kasner epoch. The time during which two same metric functions oscillates is called a Kasner era.
\subsection{In General Relativity with a massless scalar field}
The only difference with the previous case is that from now on, when the field equations are approximated by a Bianchi type $I$ model, one has $\sum_{i=1}^3p_i=1$ but $\sum_{i=1}^3p_i^2=1-\xi^2(\omega+3/2)$ where $\xi$ is a constant. Consequently, when $\omega>-3/2$, all the exponents $p_i$ can be positive. Then the metric functions are all decreasing and the BKL approximation remains valid forever: the oscillations stop. The energy density of the scalar field is positive and it respects the weak energy condition. It is the case considered in the paper \cite{BelKha73}. On the contrary, when $\omega<-3/2$, the exponents are never all positive. Consequently, the BKL approximation is not stable. It is invalid after a finite time. In this case, the energy density is negative and the weak energy condition is not respected.


\begin{thebibliography}{10}

\bibitem{Dir37}
Paul Dirac.
\newblock {\em Nature}, 139:323, 1937.

\bibitem{Kal21}
T.~Kaluza.
\newblock Zum unitätsproblem der physik.
\newblock {\em Sitzungsber. Preuss. Akad. Wiss. Phys. mat. Klasse}, 96:69,
  1921.

\bibitem{BraDic61}
Carl~H. Brans and Robert~H. Dicke.
\newblock Mach's principle and a relativistic theory of gravitation.
\newblock {\em Phys. Rev.}, 124, 3:925--935, 1961.

\bibitem{Bra97}
Carl~H. Brans.
\newblock Gravity and the tenacious scalar field.
\newblock {\em Contribution to Festscrift volume for Englebert Schucking},
  1997.

\bibitem{Zel86}
Y~A.~B. Zel'dovich.
\newblock Cosmological field thory for observational astronomers.
\newblock {\em Sov. Sci. Rev. E Astrophys. Space Phys.}, Vol. 5:1--37, 1986.

\bibitem{MatGuzUre99}
Tonatiuh Matos, Francisco~S. Guzm{\'a}n, and L.~Arturo Une{\~n}a-L{\'o}pez.
\newblock Scalar field as dark matter in the {U}niverse.
\newblock {\em Class.Quant.Grav.}, 17:1707--1712, 1999.

\bibitem{Fay03A}
S.~Fay.
\newblock Scalar fields properties for flat galactic rotation curves.
\newblock {\em Astronomy and Astrophysics}, 413:799, 2004.

\bibitem{Spe03}
D.N. Spergel et~al.
\newblock First year wilkinson microwave anisotropy probe (wmap) observations:
  Determination of cosmological parameters.
\newblock {\em Astrophys. J.}, 148:175, 2003.

\bibitem{BelKhaLif82}
V.~A. Belinskii, I.~M. Khalatnikov, and E.~M. Lifshitz.
\newblock A general solution of the {E}instein equations with a time
  singularity.
\newblock {\em Advances in Physics}, 31, 6:639--667, 1982.

\bibitem{BelKhaLif70}
V.~A. Belinskii, I.~M. Khalatnikov, and E.~M. Lifshitz.
\newblock Oscillatory approach to a singularity in the relativistic cosmology.
\newblock {\em Advances in Physics}, 19:525--573, 1970.

\bibitem{UggElsWaiEll03}
Claes Uggla, Henk van Elst, John Wainwright, and George F.~R. Ellis.
\newblock The past attractor in inhomogeneous cosmology.
\newblock {\em gr-qc/0304002, submitted for publication to Physical Review D},
  2003.

\bibitem{TopUst99}
A.~V. Toporensky and V.~O. Ustiansky.
\newblock Dynamics of bianchi $ix$ universe with massive scalar field.
\newblock {\em Phys. Rev. D.}, 37:3406, 1988.

\bibitem{BelKha73}
V.~A. Belinskii and I.~M. Khalatnikov.
\newblock Effect of scalar and vector fields on the nature of the cosmological
  singularity.
\newblock {\em JETP}, 36, 4:591--810, 1973.

\bibitem{LehMen03}
T.~Lehner and L.~Di Menza.
\newblock Revisitation of chaos in {B}ianchi {IX} universes and in generalized
  scalar-tensor cosmologies.
\newblock {\em Chaos, Solitons and Fractals}, 16:597--611, 2003.

\bibitem{Dic62}
R.~H. Dicke.
\newblock Mach's principle and invariance under transformation of units.
\newblock {\em Phys. Rev.}, 125, 6:2163, 1962.

\bibitem{RatPee88}
B.~Ratra and P.~J.~E. Peebles.
\newblock {\em Phys. Rev. D.}, 37:3406, 1988.

\bibitem{Bin99}
P.~Binetruy.
\newblock {\em Phys.Rev..}, D60:063502, 1999.

\bibitem{ColIbaHoo97}
A.A.Coley, J.~Ib{\`a}{\~n}ez, and R.J. van~den Hoogen.
\newblock {\em J. Math. Phys.}, 38:5256, 1997.

\bibitem{KitMae92}
Y.~Kitada and M.~Maeda.
\newblock {\em Phys. Rev.}, D45:1416, 1992.

\bibitem{Fay01}
S.~Fay.
\newblock Isotropisation of {G}eneralised-{S}calar {T}ensor theory plus a
  massive scalar field in the {B}ianchi type {I} model.
\newblock {\em Class. Quantum Grav}, 18:2887--2894, 2001.

\bibitem{Fay03}
S.~Fay.
\newblock Isotropisation of {B}ianchi class {A} models with curvature for a
  minimally coupled scalar tensor theory.
\newblock {\em Class. Quantum Grav}, 20, 7, 2003.

\bibitem{Fay04A}
S.~Fay.
\newblock Isotropisation of {B}ianchi class {A} models with a minimally coupled
  scalar field and a perfect fluid.
\newblock {\em Class. Quantum Grav.}, 21, 6:1609--1621, 2004.

\bibitem{Mis69}
C.~W. Misner.
\newblock Mixmaster {U}niverse.
\newblock {\em Phys. Rev. Lett.}, 22:1071--1074, 1969.

\bibitem{DemSch96}
Jacques Demaret and Christian Scheen.
\newblock Painlevé singularity analysis of the perfect fluid bianchi type-{IX}
  relativistic cosmological model.
\newblock {\em Journal of Physics A}, 29:59--76, 1996.

\bibitem{Sch97}
Christian Scheen.
\newblock Implementation of the painlevé test for ordinary differential
  systems.
\newblock {\em Theoretical Computer Science}, 187(1,2):87--104, 1997.

\end{thebibliography}
\end{document}